\documentclass[structabstract]{aa}  
\usepackage{graphicx}
\usepackage{txfonts}
\usepackage[comma]{natbib}
\usepackage{sidecap}
\usepackage{comment}
\usepackage{color}
\usepackage{booktabs}
\usepackage{upgreek}
\usepackage{bm}

\begin{document}
\title{Near-infrared study of the stellar population of Sh2-152}
\subtitle{}
\author{S. Ram\'irez Alegr\'ia\inst{1,2} \and A. Herrero\inst{1,2} \and A. Mar\'in-Franch\inst{3,4} \and E. Puga\inst{5} \and F. Najarro\inst{5} \and \\ J.A. Acosta Pulido\inst{1,2} \and S.L. Hidalgo\inst{1,2} \and S. Sim\'on-D\'iaz\inst{1,2}}

\institute{Instituto de Astrof\'isica de Canarias, E-38205 La Laguna, Tenerife, Spain. \email{sramirez@iac.es, ahd, jap, shidalgo, ssimon} \and Departamento de Astrof\'isica, Universidad de La Laguna, E-38205 La Laguna, Tenerife, Spain. \and Centro de Estudios de F\'isica del Cosmos de Arag\'on (CEFCA), E-44001 Teruel, Spain. \email{amarin@cefca.es} \and Departamento de Astrof\'isica, Universidad Complutense de Madrid, E-38040 Madrid, Spain. \and Centro de Astrobiolog\'ia (CSIC-INTA), E-28850 Torrej\'on de Ardoz, Madrid, Spain. \email{elena@damir.iem.csic.es, najarro} \\}
\date{Received 2011 / Accepted 2011}

\abstract
{The discovery of new massive star clusters and massive stellar populations in previously known clusters in our Galaxy by
means of infrared studies has changed our view of the Milky Way from an inactive to an active star-forming machine. Within this 
scenario, we present a near-infrared spectrophotometric study of the stellar content of the compact \ion{H}{II} region Sh2-152.}
{We aim to determine the distance, extinction, age, and mass of Sh2-152, using for the first time near-infrared stellar classification 
for several sources in the region.}
{Using our near-infrared ($J$, $H$, and $K_S$) photometry and the colour--magnitude diagram for the cluster field, we selected 13 
bright stars, candidate members of the reddened cluster's main sequence, for $H$- and $K$- spectroscopy and spectral classification. This
near-infrared information was complemented with an optical spectrum of the ionizing central star to confirm its spectral nature.}
{From the 13 spectroscopically observed stars, 5 were classified as B-type, 3 as G-type, 2 were young stellar objects (YSOs), 
and 3 remained unclassified (because of the poor data quality). The cluster's extinction varies from $A_{K_S}=0.5$ to 2.6 magnitudes
($A_V=4.5$ to 24 magnitudes) and the distance is estimated to be $3.21\pm0.21$ kpc. The age of the cluster is younger than 9.4 Myr and 
the lower limit to the total mass of the cluster is $(2.45\pm0.79)\cdot10^3 {M}_{\odot}$. We compare the number of ionizing photons emitted from the
OB-type stars with the Lyman continuum photons derived from the radio observations and conclude that both quantities are consistent 
for the central region of Sh2-152. In contrast, the main ionizing source of the lower region remains unidentified.}
{}
\keywords{Infrared: stars - Stars: early-types, Hertzsprung-Russell and C-M diagrams, 
mass function - Techniques: photometric, spectroscopic.}
\maketitle

\section{Introduction}

 Thanks to infrared studies such as DENIS \citep{epchtein97}, 2MASS \citep{skrutskie06}, GLIMPSE \citep{benjamin03}, 
and UKIDSS \citep{lawrence07}, we have been able to observe regions of high visual extinction, previously 
impossible to achieve with optical instruments, and discover that our Galaxy is an active star-formation machine \citep{figer08}, 
fed by massive clusters and their massive stars.
 
  Some of these massive objects, for example the Arches-Quintuplet cluster \citep{figer99},  RSGC1 \citep{figer06,davies08}, 
RSGC2 \citep{davies07}, RSGC3 \citep{clark09,alexander09} and Alicante 8 (a.k.a. RSGC4, \citealt{negueruela10}) 
are newly discovered massive clusters; but others such as CygOB2 \citep{knodlseder00} or Westerlund1 
\citep{clark05}, are previously known clusters with a newly detected massive stellar population. This increase 
in massive cluster discoveries has been complemented with new catalogues of cluster candidates, for example 
\citet{bdb03,bica03}, \citet{dutra03}, \citet{mercer05}, and \citet{froebrich07}. If we also consider that hundreds of 
Galactic massive clusters remain unknown \citep{hansonpopescu08}, it is easy to grasp the need to generate 
projects dedicated to finding and characterizing these new massive objects.
  
 This re-evaluation of our ideas of the Milky Way motivated us to develop the MASGOMAS (MAssive Stars in 
Galactic Obscured MAssive clusterS) project. Within  this project, we have completed a photometric near-infrared 
broad band ($J$, $H$ and $K_S$) catalogue of 44 cluster candidates extracted from previous catalogues 
\citep{dutra01,bdb03,bica03} and we have continued with a medium resolution multi-object spectroscopic 
follow-up for 9 of the most promising candidates. A detailed description of MASGOMAS is given in the first paper 
of the project \citep{marin09}.

\begin{figure} 
\centering
\includegraphics[width=3.7in, angle=0]{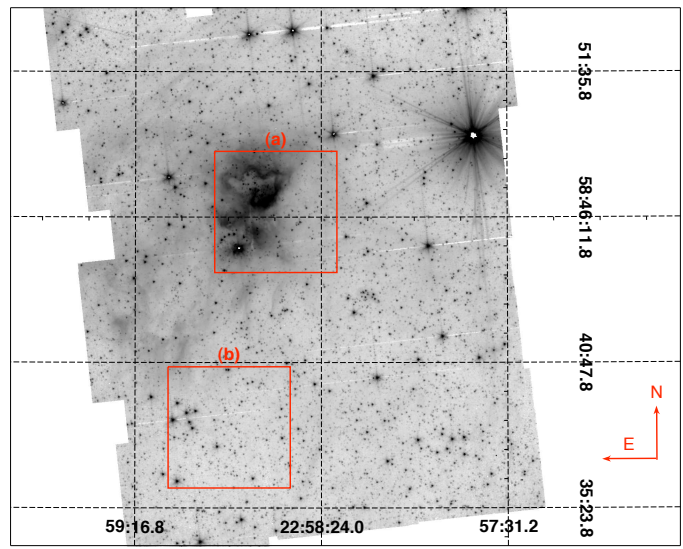}  
            \caption{Spitzer $3.6\,\mu$m image. 
It covers the target (a) and control (b) fields for Sh2-152. Red squares are 
equivalent to the LIRIS field of view (4.2 arcminutes length).}
       \label{id436_spitzer}
\end{figure}

  One of our candidates is the star-forming compact \ion{H}{II} region Sh2-152 
(${\alpha}_{2000}=22^{\mathrm {h}}58^{\mathrm {m}}45^{\mathrm {s}}$ and ${\delta}_{2000}=+58\degr46\arcmin50\arcsec$).
This region is located in the Perseus arm, close to the Galactic plane ($l=108\,\fdg76,
b=-0\,\fdg95$) and stands out from the background in the Spitzer 3.6\,$\mu$m
image  shown in Fig. \ref{id436_spitzer}. Sh2-152 contains
two IRAS sources: IRAS 22566+5830 and IRAS 22566+5828 \citep{kleinmann86}. 
The first is located in the central zone of the cluster and would be 
excited by an O9\,V star \citep{crampton78}, labelled by \cite{russeil07} 
as star number 4, and the second IRAS region, located $\sim2$ arcmin SE from 
the central section, appears more extended and less populated than the 
central IRAS 22566+5830.
  
  In the Sh2-152 field of view we find five masers, all of them presented in 
the false-colour LIRIS image (see Fig. \ref{mos_stars}): one 6.7 GHz methanol \citep{szymczak00},
three water \citep{palagi93,harju98}, and one hydroxyl (OH) \citep{wouterloothabing85}.
  These masers are indicative of star formation and in the case of the hydroxyl one, of high-mass star
 formation \citep{zinnecker07}. Its presence makes this cluster an interesting candidate as a 
still young and star-forming active region. Following this idea, \citet{chen09} used $K'$, and $\rm H_2$ photometry to
detect $\rm H_2$ emission in Sh2-152, implying that the cluster hosts ongoing star formation.
 
 However, even if this region has been the object of near-infrared observations its stellar content has not yet
been characterized spectroscopically. Only the ionizing central star has been optically observed and spectroscopically
classified \citep{crampton78,russeil07}.

   Another interesting point about Sh2-152 is its distance and extension.
\citet{crampton78}, reported 3.6 kpc as the first estimate of the cluster's
distance using the spectroscopic parallax to the ionizing star. Later, in a series of studies 
observing $\rm H_2O$ masers, NH${}_{3}$ emission lines, and both 
\element[][12]{CO}($J=2\rightarrow1)$ and \element[][12]{CO}($J=3\rightarrow2)$ transitions, \citet{wouterloot86}
 estimated the  distance to Sh2-152. Assuming that Sh2-152
is located in the Perseus arm, between 2 and 6 kpc or an average distance of
3500 pc, they confirm this conjecture first with the $\rm H_2O$ maser observation,
 and then, with the ammonia core \citep{wouterloot88}. 
In any case, in the third work of this series, \citet{wouterloot93}, determined 
a (kinematic) distance of 5.3 kpc through radial velocity measurements of 
\element[][12]{CO}($J=2\rightarrow1)$ and \element[][12]{CO}($J=3\rightarrow2)$ lines 
for IRAS 22566+5830. In the same year, \citet{harju93} presented consistent results, 
based on observations of ammonia clumps, with the initial assumption of a distance 
of 3.5 kpc for Sh2-152. 

 On the basis of long-slit optical spectroscopy for the 
central region of Sh2-152, \citet{russeil07} then calculated a distance of 2.39 kpc. This distance was derived 
from the spectral classification of the ionizing central star (the aforementioned 
star number 4), although we note that the position of the star is inaccurate, probably because of the
 spatial resolution of the associated images. 

 The cluster extension has been estimated considering only visual or 
infrared images, but from an inspection of Spitzer $3.6, 4.5, 5.8$, and $8.0\,\mu$m images it
 is easy to see that the cloud around the cluster is more extended. Spectroscopy of stellar objects, covering
 not only the central region of Sh2-152 but the area marked with the (a) square (see Fig. \ref{id436_spitzer}), would allow
 us to determine their individual distances and the extension of the cluster. 

  With the mass of associated molecular clouds and high-mass protostellar objects (HMPO), 
 the mass of Sh2-152 can be indirectly estimated. \citet{ao04} determined the dust mass
 of the CS, \element[][13]{CO}, and C\element[][18]O cores, mapping their respective lines. Assuming
 a distance of 3.5 kpc, from the column densities of the three observed molecular lines, they 
measured masses of  $1.78\cdot10^3 {M}_{\odot}$ (CS core), 
$9.20\cdot10^3 {M}_{\odot}$ (\element[][13]{CO} core), $0.29\cdot10^4{M}_{\odot}$, and 
$1.20\cdot10^4 {M}_{\odot}$ (for two C\element[][18]O cores).

 From the CO column density and adopting a distance of 5.3 kpc \citep{wouterloot93}, \citet{guan08} 
determined a mass of $3.81\cdot10^4 M_{\odot}$ for the associated CO core, from which they observed 
the emergence of mass outflows. 
The size of the CO cloud determined by them was 4.1 pc, similar to the 4.3 pc reported by
 \citet{ao04} to the \element[][13]{CO} core, but larger than the other three observed 
cores (1.8 pc for the CS core and 1.7-3.0 pc for the C\element[][18]O ones). Using 
850\,$\mu$m polarimetry, \citet{curran04} studied the region. For IRAS 22566+5828,  
they found outflows and a high-mass protostellar object (HMPO) with an estimated 
total mass (dust) of 3500 $M_{\odot}$. In the case of IRAS 22566+5830, they 
detected a magnetic field aligned north--south. In their work, they 
used 5.0 kpc as the distance to the object and 100 as the gas-to-dust ratio.
 
  As the physical parameters depend on the square of the distance,
 it is crucial to have a reliable determination of the distance. We therefore refine the distance and extension of
 Sh2-152 using near-infrared photometry for the field and spectra for 13 stars. This would provide the 
first distance and extension estimates using more than one spectrum for candidates 
of the cluster stellar population. Likewise, we plan to obtain the total cluster
 mass by using its stellar population and luminosity function.
 
  This paper is organized as follows. Section 2 details the observations, 
as well as the related data reduction procedures. The 
analyses of the photometry and spectroscopy (spectral classification) are described
in Sect. 3. The cluster parameters are discussed in Sect. 4. Finally, Sect. 5 
summarizes our conclusions.

\section{Observations}

 Our study is based on broad$-$band infrared imaging ($J$, $H$, $K_S$) and medium 
resolution multi-object spectroscopy ($H$ and $K$), acquired with LIRIS, a 
near-infrared imager/spectrograph mounted at the Cassegrain focus of the 4.2 m William 
Herschel Telescope (Roque de Los Muchachos Observatory, La Palma). 
 
 This infrared data was complemented with optical spectroscopy for the central ionizing source of the cluster,
 obtained with FIES at the Nordic Optical Telescope (NOT, Roque de los Muchachos 
Observatory, La Palma). A summary of the observations is given in Table \ref{tabla_obs}.

\begin{table}
\caption{Summary of imaging and spectroscopic observations.}

\begin{center}
\begin{tabular}{ccccc}
\toprule
RA (J2000) & Dec (J2000) & Filter & Exp. time & Seeing \\
\,[$\mathrm{{~~}^{h}{~~}^{m}{~~}^{s}}$] & [$\mathrm{{~~}^{\circ}~~'~~''}$] & & [s] & [$''$]\\

\midrule
 \multicolumn{5}{l}{Sh2-152 field imaging:} \\
\midrule
 22 58 37	&  +58 46 24  &   $J$      & 222.30   &  0.62\\
 		&  		      &   $H$      & 93.12      &  0.69\\
  		&  		      &   $K_S$  & 66.12     &  0.58\\
\midrule
\multicolumn{5}{l}{Multi-object infrared spectroscopy ($R\sim2500$):} \\
\midrule
 22 58 45	&  +58 46 50	  &  $H$  &   2400   &  1.04    \\
                   &                           &  $K$  &   2400   &  0.98 \\
\midrule
\multicolumn{5}{l}{Optical spectroscopy ($R\sim46000$):} \\
\midrule
 22 58 40.8    &  +58 46 58.2     &   370-730 nm & 1555 & 1.50    \\
\midrule
\multicolumn{5}{l}{Control field imaging:} \\
\midrule
 22 58 50	&  +58 38 24	  &	$J$      & 79.34      &     0.62\\
 		&  	  	&	        $H$	  & 79.34	&	0.59\\
  		&  		&	       $K_S$  & 79.34      &     0.65\\
\bottomrule
\end{tabular}
\end{center}
\label{tabla_obs}
\end{table}

\subsection{Imaging}
 
\begin{figure*} 
\centering
\includegraphics[width=6.0in]{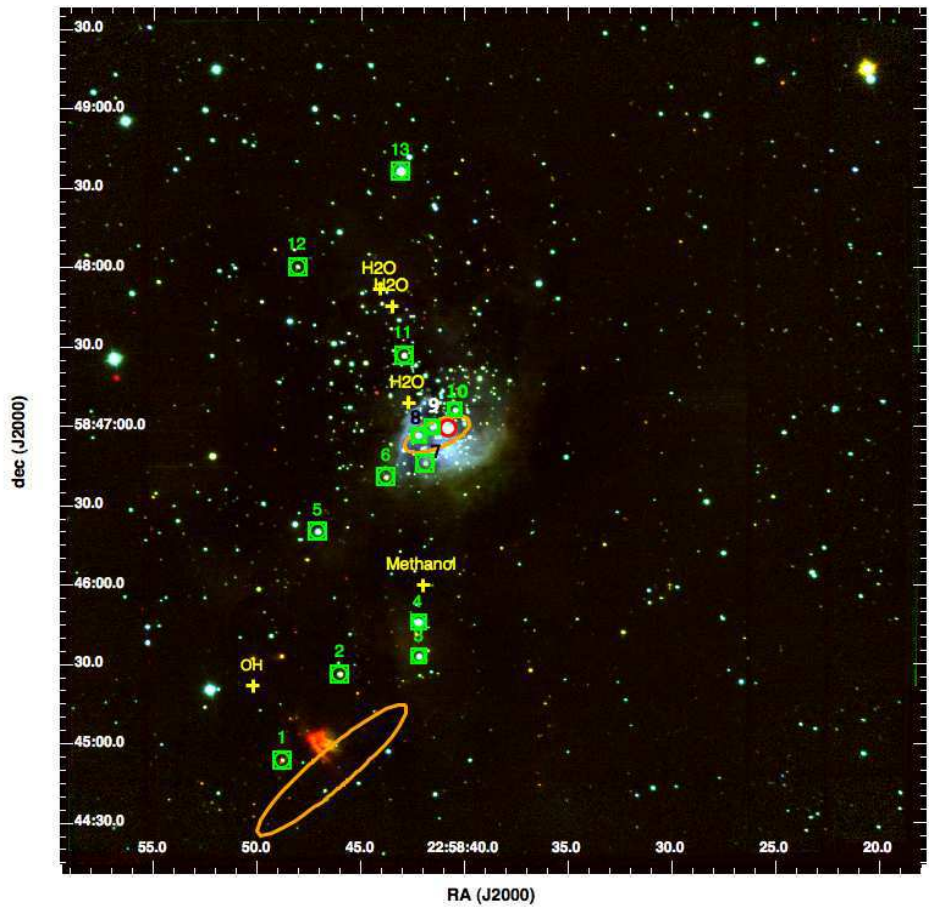}  
            \caption{LIRIS false-colour image of Sh2-152 (blue=$J$, green=$H$, 
red=$K_S$); marked with green squares, the stars observed spectroscopically in 
the near-infrared. In the centre, indicated by a red circle, the central star observed 
spectroscopically in visual. The positions of 
the masers within the field are indicated by yellow crosses. The (orange) ellipses correspond to the 1-$\sigma$ 
position uncertainties for the IRAS 22566+5830 (central) and IRAS 22566+5828 (lower) sources. Black and white
colours for central stars are used for visibility reasons.}
       \label{mos_stars}
    \end{figure*}

 The cluster candidate images were obtained on 2006 July 22 and the control 
field was observed on 2009 November 27. In both runs, seeing was between $0.58\arcsec$ and $0.69\arcsec$.
Observations were carried out using LIRIS, an infrared camera equipped with a 
Hawaii $1024\times1024$ HgCdTe array
detector, with a field of view of $4.2\arcmin \times4.2\arcmin$ and a spatial scale of 
$0.25\arcsec{pixel}^{-1}$, using $J$
(${\lambda}_C =1{.}250\,\mu$m, $\Delta\lambda= 0{.}160\,\mu$m), $H$ 
(${\lambda}_C =1{.}635\,\mu$m, $\Delta\lambda= 0{.}290\,\mu$m), 
 and ${K}_{S}$ (${\lambda}_C =2{.}150\,\mu$m, $\Delta\lambda= 0{.}320\,\mu$m) 
filters. To improve cosmic-rays and bad-pixel rejection, and to construct the sky image for the 
sky subtraction, we observed in eight-point dithering mode. Data reduction (bad pixel mask, flat correction, sky
subtraction, and alignment) was done with FATBOY \citep{eikenberry06} and geometrical distortions
were corrected with the LIRIS reduction package, LIRISDR\footnote{http://www.iac.es/project/LIRIS}. The final 
false-colour image for Sh2-152 field is shown in Fig. \ref{mos_stars}.

 The instrumental photometry was done with DAOPHOT II, ALLSTAR, and ALLFRAME 
\citep{stetson94}, and was cleaned of
non-stellar and poorly measured objects based on two parameters. The first one is the
sharpness index, which is the difference between the observed width of the object and the PSF 
width. Objects such as hot pixels have a negative sharpness values while a ``broad''
object, such as galaxies or pairs of unresolved stars, have a positive sharpness. The sharpness was complemented
with the estimated standard error in the measured magnitude, $\sigma$. Using values of sharpness
in the range $-0.25<sharp<0.25$, and $\sigma < 0.1$, we filtered 20\% of the detections, all
of them with $K_S$ over 16.

The photometric calibration (including a colour term) was done using the 2MASS
catalogue \citep{skrutskie06} for the three filters. We selected isolated and 
non-saturated stars, with photometry in 2MASS well-distributed over our whole 
image and the colour--magnitude diagram. For the stars with $K_S$ brighter than
11, we adopted the 2MASS photometry, owing to their image saturation.

 The astrometric calibration was performed with SKYCAT, matching physical
 and equatorial coordinates for 21 sources, well-distributed in our LIRIS image. 
Typical values of the RMS of the fitting were smaller than 0.15 arcsec for the three 
bands, which is adequate for the mask design.

\subsection{Infrared spectroscopy}
 The infrared spectroscopic observations were obtained on 2008 August 19 with seeing 
$\sim1\arcsec$ and using the multi-object spectroscopic mode (MOS) and
 the HK pseudogrism, with a resolution of $\lambda/\Delta\lambda \sim 2500$. We observed two 
A0\,V stars as telluric standards: HD 223386 for the $H$ pseudogrism observations and Hip 107555, 
for the $K$ one.
  The mask contained 13 slits of width 0.8 arcsec and length 6 arcsec and was centred
 at ${\alpha}_{2000} = 22^{h}58^{m}45^{s}$ and 
${\delta}_{2000} = +58\degr46\arcmin50\arcsec$. In Fig. 
\ref{mos_stars}, we mark the stars observed spectroscopically in the near-IR.
 The selection criteria for multi-object spectrocopy is detailed in section \ref{sec:cmd}.
  
  We observed using an ABBA strategy (a star is located in positions A and B, in the slit, and then sequentially 
changed); with this mode, we are able to remove the sky from the spectra by simply subtracting 
spectra at position A from spectra at position B, and vice versa. Flat-fielding, tracing, 
sky subtraction, coaddition, and extraction  were done using LIRISDR, a package 
developed specifically for LIRIS data that uses the information from the mask design files. 
Thanks to the combination of the individual spectra (eight for each band), we discarded 
cosmic rays and hot pixels that could mimic spectral lines. For wavelength calibration,
argon and xenon lamps were observed, using both lamps (continuum subtracted) to 
calibrate the $K$-band spectra and the argon lamp for $H$-band calibration.
  
  Telluric subtraction was done using XTELLCOR \citep{vacca03}, an IDL program that, applying
a high$-$resolution synthetic model of an A0\,V star (the spectral type of our standard) over the 
observed telluric standards, produces the calibration spectrum with the telluric lines. This
spectrum was then used to correct our science spectra, with the IRAF\footnote{IRAF is distributed 
by the National Optical Astronomy Observatories, which are operated by the Association of 
Universities for Research in Astronomy, Inc., under cooperative agreement with the National 
Science Foundation.} task TELLURIC.
 
\subsection{Optical spectroscopy}
 The optical spectra for the central ionizing source of Sh152-2 (star number 4 from \citealp{russeil07})
 was obtained on 2009 November 12, using the FIES cross$-$disperser high$-$resolution echelle 
 spectrograph, at the 2.5 m  NOT. We used FIES in medium resolution mode $(R=46000)$, with a 
 fibre size of $1.3\arcsec$ and the spectral range of 370 to 730 nm covered in a single fixed setting. 
 The exposure time for the individual spectra was 25.83 minutes, and the seeing during the observations 
 was 1.5 arcseconds.

\section{Results \label{res}}
  Previous near-infrared studies of Sh2-152 were based on the 2MASS photometry, thus their results  were
  limited to magnitude $K_S\sim13$. With our new observations, we obtained near-infrared photometry with 
  limiting magnitude $K_S\sim20$ for the cluster and control field. These data were used to select the OB-type
  star candidates from the cluster main-sequence for the spectroscopic follow-up. 
 
\subsection{Colour--magnitude diagram}
\label{sec:cmd}
 
\begin{figure*}
\centering
\includegraphics[width=5.5in]{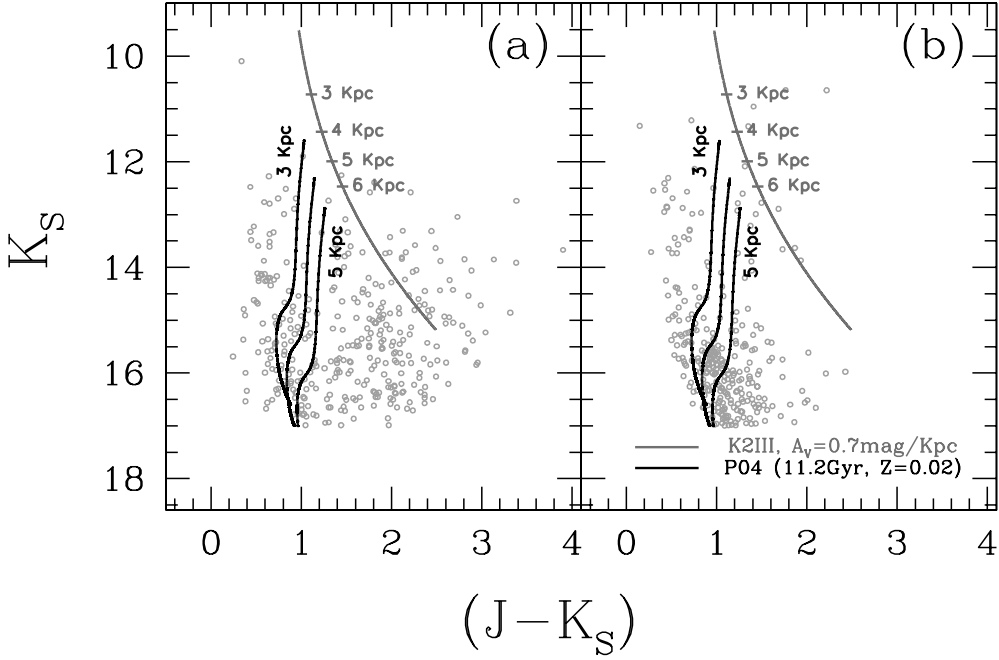} 
            \caption{Calibrated colour--magnitude diagrams for Sh2-152 (a) and the
 control field (b). The black lines represent  isochrones of solar metallicity and 
11.2 Gyr, located at three distances (3.0, 4.0 and 5.0 kpc) and the grey curve, the 
position of a K2\,III star at several distances. The isochrone was generated 
using the \citet{pietrinferni04} stellar evolutionary library, the \citet{castelli03} 
bolometric correction library and the \citet{rieke89} extinction law with $R=3.09$ 
\citep{rieke85}. For the K2\,III sequence, $M_V$ was obtained from \citet{cox00} and 
IR intrinsic colours from \citet{ducati01}.}
       \label{id436control_CMDs}
    \end{figure*}

 The cluster candidate and control field colour--magnitude diagrams are shown in 
Fig. \ref{id436control_CMDs}. In these diagrams we see how the cluster's main 
sequence is scattered following the reddening vector due to the differential reddening 
over the field. This scatter also affects part of the local dwarf sequence, which is the
 principal feature in the control field CMD.
  Because of the differential extinction over the candidate field, 
cluster stars have redder $(J-K_S)$ colours and dimmer $K_S$ magnitudes.
  
 The selection of OB-type candidates considers this effect by selecting stars 
with $(J-K_S)$ in the range $0.5-4.5$ and $K_S\textless 13$. This first requirement would select 
stars embedded in the cluster's natal cloud or their own particular natal cloud, while the second 
requirement comes from the instrumental limitation of spectra of high signal-to-noise ($SNR$).
 It is necessary to keep in mind that telluric variations impose an upper limit on the infrared exposure times.
 
  For the candidate selection, we also took into account the positions of these stars over 
the field of view. First of all, the cluster's boundaries do not appear to be clearly defined in
 the LIRIS near-IR images. Including stars that are not only in the central region of the image 
would help us to derive information about the membership of stars separated from the central 
part of Sh2-152. Secondly, the number and positions of slits in the mask is limited by 
spectral overlap in the dispersion axis and the spectral features that we wish to observe.
   
  From the target and control field CMDs, it is also possible to estimate roughly the 
cluster candidate distance. Observing both CMDs, we can identify the local dwarf sequence, 
formed by dwarf stars situated in the solar neighbourhood, as a vertical strip
 between $(J-K_S)=0.4$ and $(J-K_S)=1.0$. This strip looks sharper for the control field 
CMD in the whole $K_S$ magnitude range but, in the target CMD, this sequence spreads for 
$K_S \textgreater 15$. The effect can be understood as the extra reddening caused by
 Sh2-152 over the disc stars located behind the cluster candidate. Hence, any star from 
the local sequence brighter than the magnitude where this sequence widens would be 
located between the cluster candidate and us. By locating this point, we can make a 
first estimation of the distance to Sh2-152.
 
 A method of obtaining this estimate is presented in \citet{marin09}. Considering 
$11.2\pm0.75$ Gyr as the characteristic age for stars in the solar neighbourhood
 \citep{binney00}, in their CMDs they plot isochrones with this age at different 
distances and thus under different extinctions. They then locate the point where
 target and control field local sequences start to differentiate into this set of
 isochrones. The closest isochrone to this point indicates a rough distance 
estimate for the cluster. 
 
 After including three isochrones in our CMDs for 3.0, 4.0, and 5.0 kpc, we can see
that differences appear between the 3.0 and 4.0 kpc isochrones (Fig. \ref{id436control_CMDs}).
Therefore, a first estimate of the distance of Sh2-152, completely independent of the 
spectrophotometric distance estimate, would be less than 4.0 kpc.
 
\subsection{Completeness tests}

\begin{figure}
\centering
\includegraphics[width=3.3in]{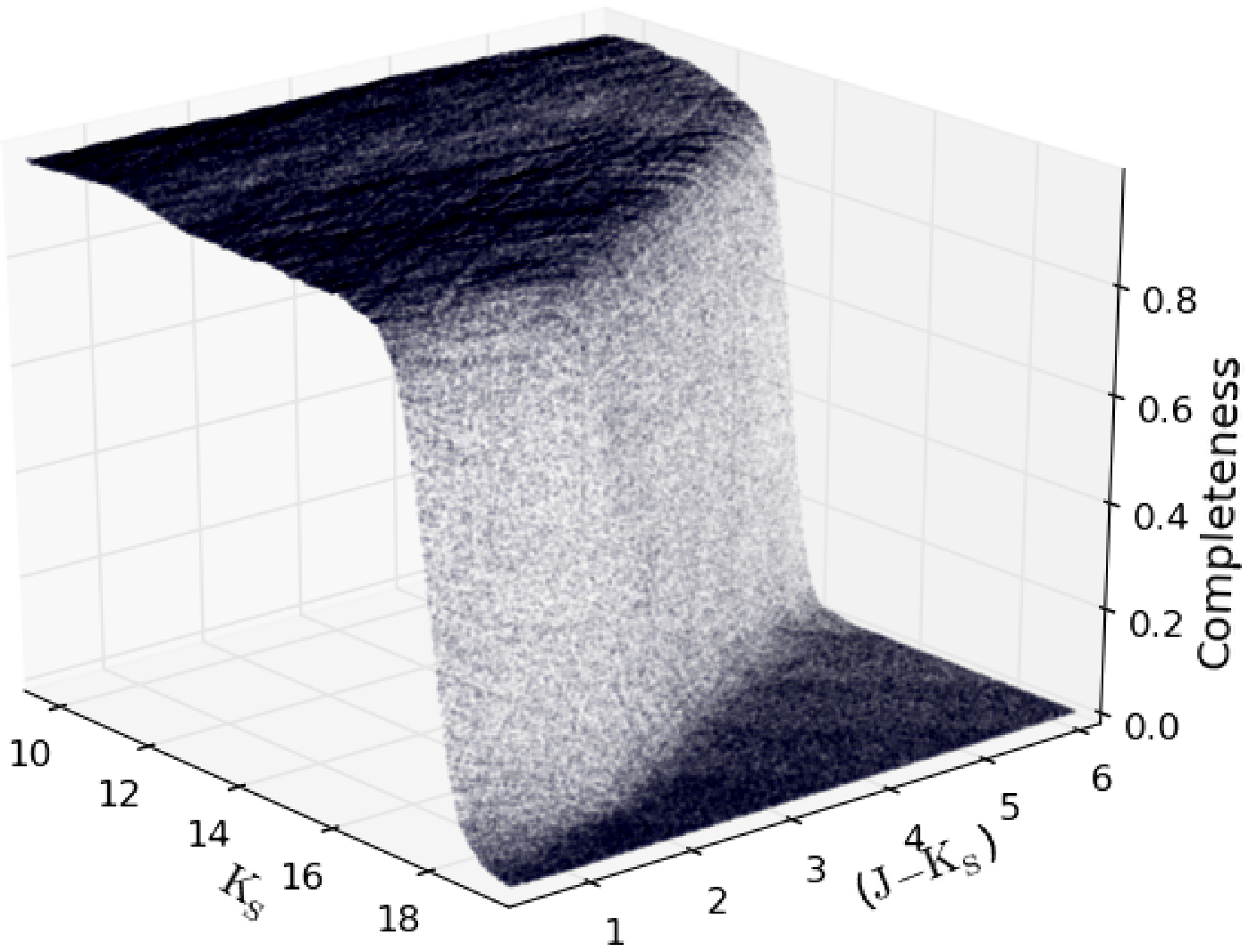}\\
\includegraphics[width=3.3in]{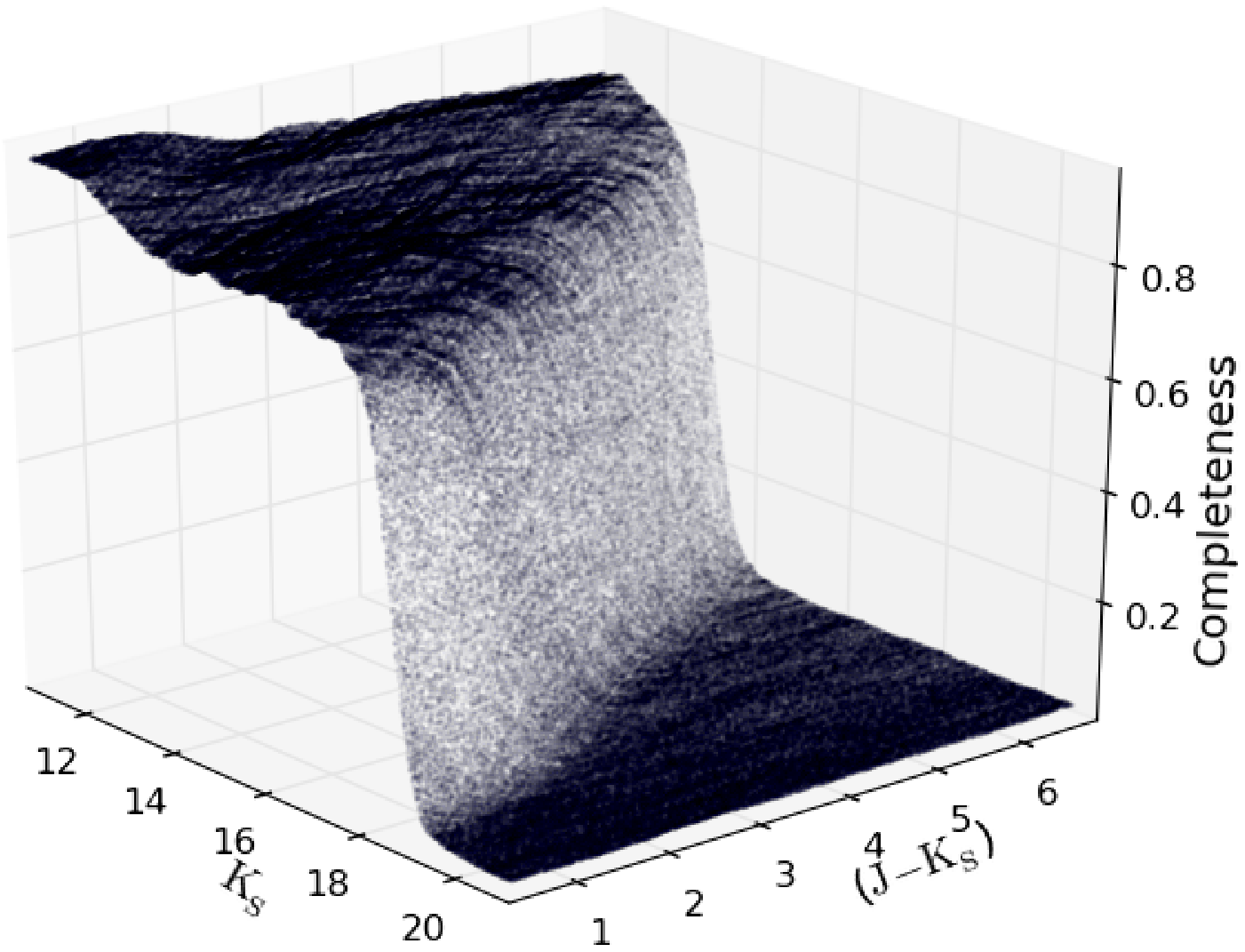}
\caption{Surface formed by the points in colour--magnitude--completeness space is shown. The top figure
corresponds to the Sh2-152 field and the bottom figure, to the control field. Completeness ranges from 0 to 1.}
       \label{completeness}
    \end{figure}

 To precisely remove the disc stellar population from the cluster's 
luminosity function, and quantify the instrumental sensitivity to stellar 
magnitudes and colours, we carried out a completeness test, as described in 
\citet{hidalgo08} and \citet{aparicio95}.

 Completeness tests consist of adding artificial stars to the images and then, 
repeating the photometry using the same parameters and procedures as for target 
and control field photometry. For our tests, we included 480200 artificial stars 
over 200 images (for $J$ and $K_S$ filters) with $K_S$ instrumental magnitudes 
between 12 and 22 and colour ($J-K_S$) between 0 and 6; with these limits, we 
homogeneously covered the whole range of the CMDs. To avoid the creation of extra 
crowding, the stars were added following a network, within the full range of 
coordinates into the image, and using a star-to-star distance of two times the 
PSF radius plus one. In this array, star positions are fixed from image to image, 
but $J$ and $K_S$ magnitudes vary. 
 
 Finally, for each artificial star of magnitude $K_{S,1}$ and colour $(J-K_S)_1$, 
 we defined a section of $K_{S,1}\pm0.4$ and $(J-K_S)_1\pm0.4$, where we 
counted the number of recovered and injected stars. The ratio of these numbers is 
the completeness index assigned to that star. The completeness test reproduces
the detection limit for the CMD and gives a completeness of more than 0.90 for
 over 97\% of the CMD. The value of the completeness factor according to the 
position in the CMD is shown in Fig. \ref{completeness}. It can be seen in the top diagram
that for $K_S$ fainter than 16.5, the completeness drops to 0.8, 
meaning that 80\% of the stars would be detected at that magnitude. 

\subsection{Near-infrared spectral classification}

 \begin{figure}
\centering
\includegraphics[width=3.5in]{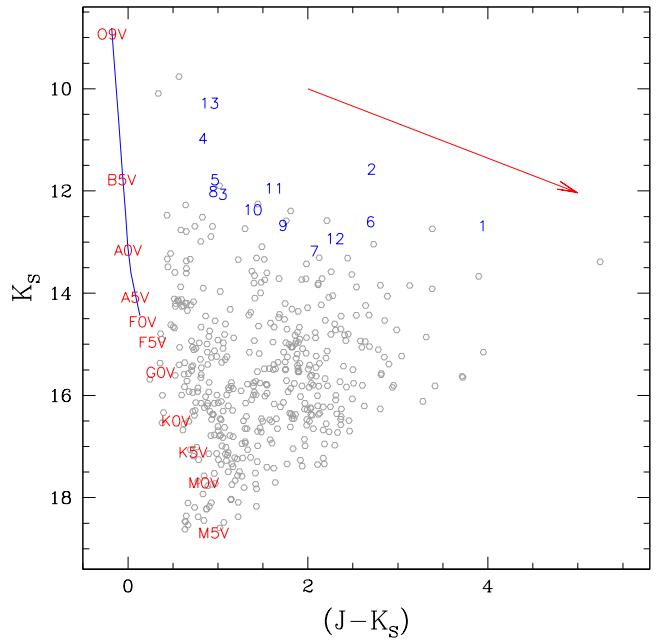}
     \caption{Calibrated colour--magnitude diagram for Sh2-152. The red segmented
 arrow shows $A_V=20$ mag ($A_{Ks}=2.04$) and the main sequence is located at the 
distance determined in this study (3.21 kpc). The continuous blue line in the upper part 
of the main sequence corresponds to the best fit made to project the CMD stars to 
the main sequence, following the reddening vector. The position of the spectroscopically 
observed stars is marked with blue numbers.}
       \label{id436_CMDs}
    \end{figure}
    
\begin{figure}
\centering
\includegraphics[width=3.5in]{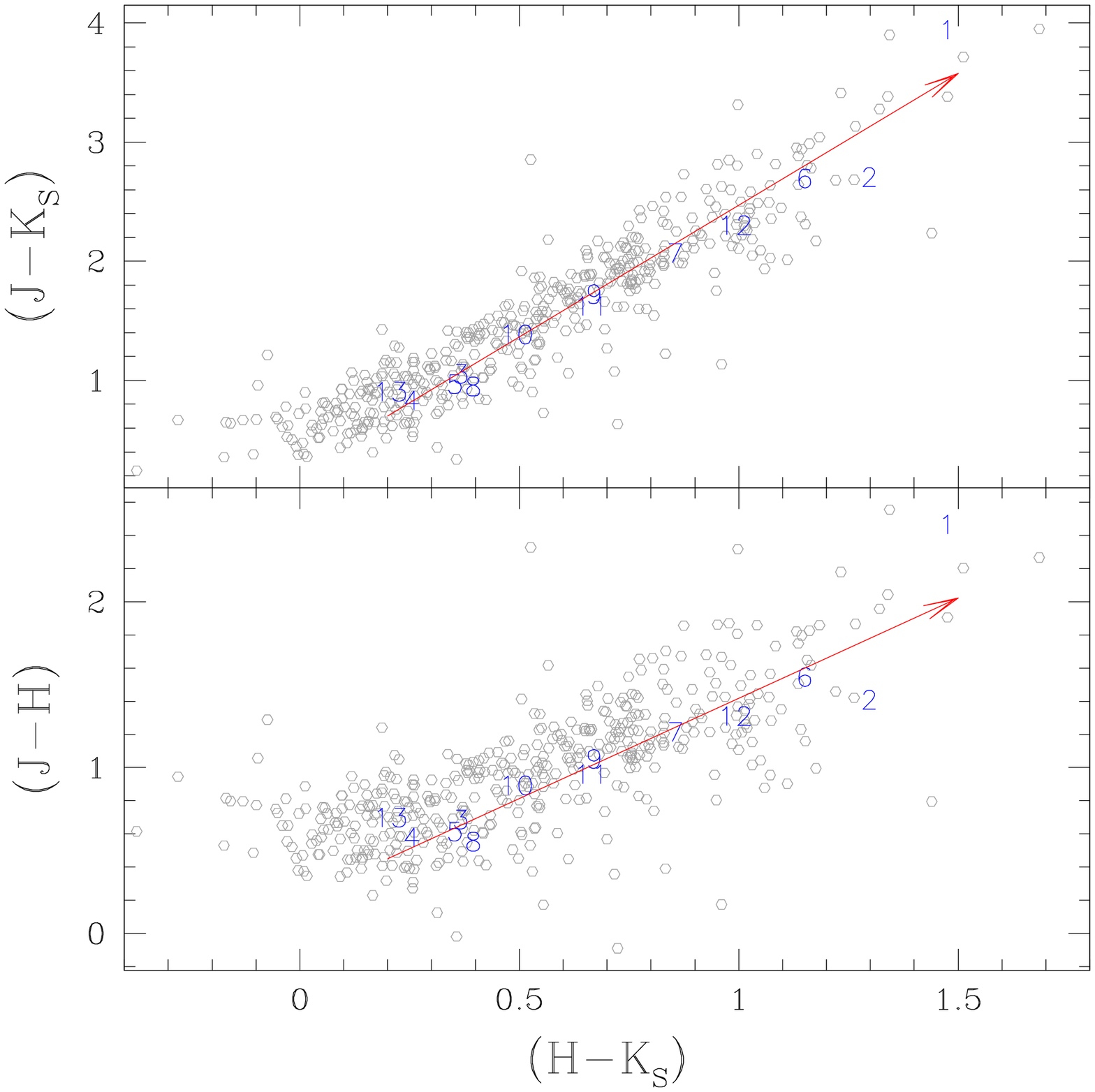}
 \caption{Calibrated colour--colour diagrams for Sh2-152. The blue numbers
 corresponds to the spectroscopically observed stars. Red arrows in both diagrams
 show the reddening vector.}
       \label{id436_CCs}
    \end{figure}

 We based our near-IR spectral classifications for the OB stellar types on those of \citet{hanson96} for 
the $K$-band and \citet{hanson98} for the $H$-band, and for the later spectral 
types on those of \citet{meyer98} and \citet{wallacehinkle97}. The characteristic lines
for each spectral type were complemented 
with a visual comparison between our spectra and other spectral catalogues,
 with the $H$- and $K$-bands at similar resolutions \citep{ivanov04, ranade04, 
ranade07, hanson05}. Table \ref{data_stars} contains the coordinates, near-infrared
 magnitudes, and spectral types of the spectroscopically observed 
stars. The positions in the CMD and CC diagrams for the observed stars are
 presented in Figs. \ref{id436_CMDs} and \ref{id436_CCs}.
 
  Spectral classification in the $K$-band can be problematic. The main feature in 
this band for late B to early F-type stars is Brackett $\gamma$ at 2.166\,$\mu$m, 
but this line is adjusted in the A0\,V modelling when constructing the telluric-line
correction spectra. Any over- or under-correction could therefore add noise 
to this particular line. Complementing the analysis with the $H$-band spectral 
information overcomes this problem.
 
  Another consideration stems from the nebular differential emission that dominates the central 
zone of the cluster. This affects the stellar spectra in two ways. First, the emission can 
contaminate the spectra, making it hard to distinguish the nature of the lines, whether stellar
 or nebular. This is noticeable in star \#8, whose wide Brackett-series emission lines seem to be
 contaminated with a narrow absorption line. This absorption line is an artefact resulting from
 the nebular contribution included in the slit and reversed through the sky subtraction.

\begin{figure*} 
\centering
\includegraphics[width=5.5in]{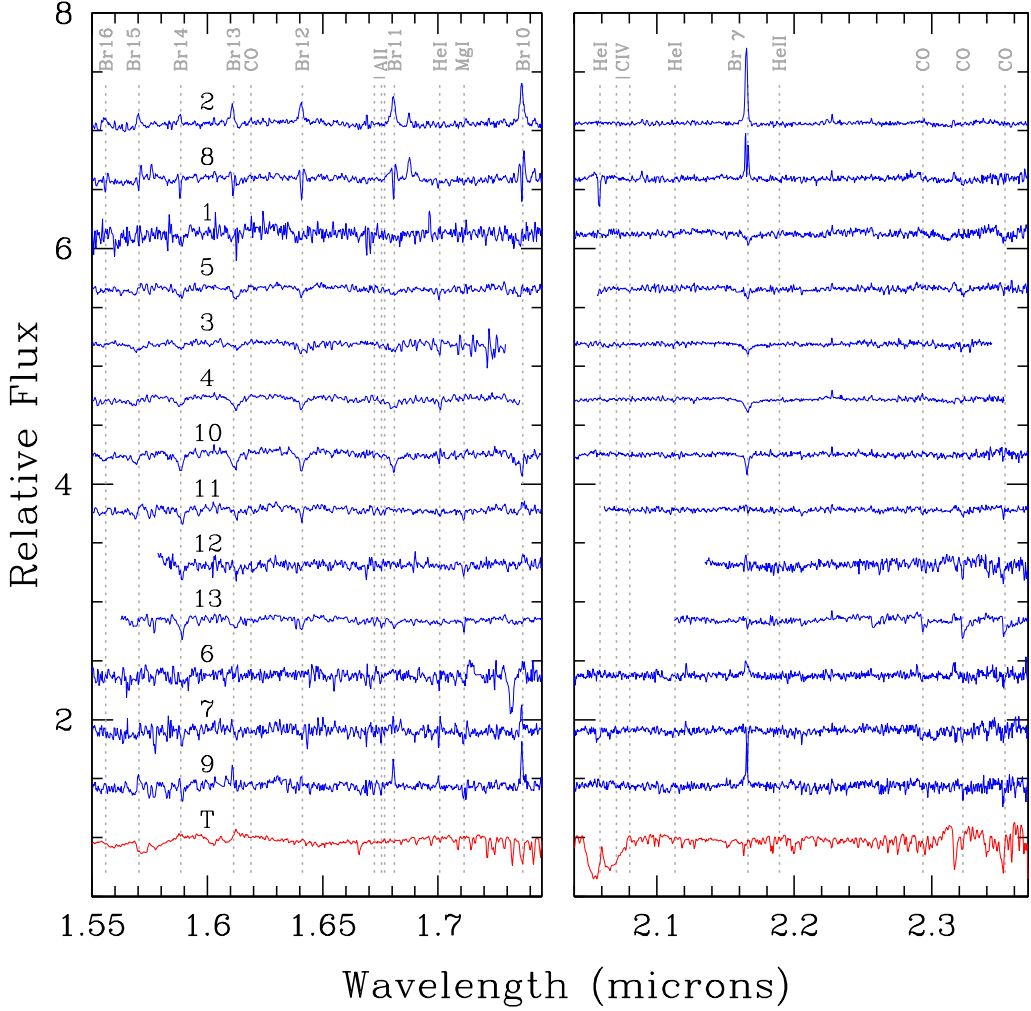}  
            \caption{Individual spectra for the observed Sh2-152 stars. The spectra are arranged, 
from top to bottom, into YSOs, early types, late types, and those without spectral classification. The 
final red spectrum corresponds to the telluric correction.}
       \label{spectra_nearIR}
    \end{figure*}

   The second problem is that, owing to the inhomogeneous nebular distribution, for the same 
slit we can have sky with an added nebular contribution relative to the local sky or even no
nebular contribution to the star position. This causes an artificial subtraction of nebular lines
or, in some cases, an artificial emission line. This effect is easy to detect through a visual 
inspection of the individual A-B or B-A spectra, detecting lines that pass from absorption to
emission as we change from an A-B to a B-A image. Examples of spectra with this behaviour are 
numbers 7 and 10, both located in the central region of the cluster, where inhomogeneous 
nebular emission is dominant. The final near-IR spectra are shown in Fig. \ref{spectra_nearIR}.
 
\begin{table*}
\caption{Spectroscopically observed stars.}
\begin{center}
\begin{tabular}{ccccccc}
\toprule
 ID & RA (J2000) & Dec (J2000) & $J$ & $H$ & $K_S$ & Spectral type\\
  & [$\mathrm{{~~}^{h}{~~}^{m}{~~}^{s}}$] & [$\mathrm{{~~}^{\circ}~~'~~''}$] & [mag] & [mag] & [mag] &  \\
 \midrule
1  & 22 58 48.774 & +58 44 53.72 & 16.598 & 14.158 & 12.745 & B1 V \\
2  & 22 58 46.007 & +58 45 26.23 & 14.258 & 12.870 & 11.633 & YSO \\
3  & 22 58 42.189 & +58 45 32.88 & 13.131 & 12.434 & 12.127 & B2--3 V\\
4  & 22 58 42.221 & +58 45 45.68 & 11.821 & 11.229 & 11.034 & B2--3 V\\
5  & 22 58 47.081 & +58 46 20.00 & 12.758 & 12.134 & 11.840 & B2 V\\
6  & 22 58 43.785 & +58 46 40.59 & 15.284 & 13.754 & 12.665 & $\cdots$ \\
7  & 22 58 41.859 & +58 46 45.99 & 15.242 & 14.035 & 13.240 & $\cdots$ \\
8  & 22 58 42.215 & +58 46 56.47 & 12.967 & 12.407 & 12.071 & YSO\\
9  & 22 58 41.541 & +58 46 59.56 & 14.395 & 13.343 & 12.732 & $\cdots$ \\
10 & 22 58 40.422 & +58 47 06.06 & 13.762 & 12.863 & 12.430 & B5 V\\
11 & 22 58 42.920 & +58 47 26.67 & 13.572 & 12.614 & 12.010 & G6.5 V\\
12 & 22 58 48.062 & +58 48 00.22 & 15.221 & 13.923 & 12.992 & G8 V\\
13 & 22 58 43.076 & +58 48 36.34 & 11.203 & 10.491 & 10.344 & G8-9 III\\
\bottomrule
\end{tabular}
\end{center}
\label{data_stars}
\end{table*}
 
Star \#1 is located in a very interesting sector of the field. Its surroundings look to be clear
in the $J$-band but for $K_S$ a bright arc and a nebulosity arises. This arc even rivals 
the central zone of Sh2-152 in brightness (Spitzer ch-4, 8.0\,$\mu$m). Because of this 
nebulosity, star \#1 is the most reddened star in our set but we cannot determine whether 
it is the ionizing star of the IRAS 22566+5828. The star was classified as  B1\,V, based on
the shape and depth of its Br$\gamma$ line and the presence of helium (2.11\,$\mu$m). 
The $H$-band is quite noisy, and the \ion{He}{I} 1.70\,$\mu$m is contaminated by nebular
emission. Owing to this noise, produced by its lower position in the detector causing the 
lost of half of the flux, the stellar type could easily be two subtypes earlier or later.
 
 Spectrum \#2 is characterized by its strong emission Brackett lines. This object is located below the
main-sequence line in the colour--colour diagrams (Fig. \ref{id436_CCs}). Its near-IR colour excess, which 
is characteristic of young stellar objects, and the dominating
 emission lines over the whole spectra lead us to classify it as a YSO. 
 
   Stars \#3 and \#4 have similar features and their spectral types vary in the range B2-3\,V 
according to their $H$-band spectra. Star \#3 spectra is noisier than star \#4, and this explain
the weaker Brackett series. However for both spectra the series are present until Br15 (1.57\,$\mu$m), 
confirming the similar spectral types of both stars. We also detect for both stellar spectra 
the \ion{He}{I} 1.70\,$\mu$m line. The $K$-band  spectrum does not clearly display the \ion{He}{I} 2.11\,$\mu$m
 line, but the Br$\gamma$ line is again consistent with a B2-3\,V spectral classification.

 For star \#5, the spectral type is difficult to define. The spectrum clearly shows the lines Br12,
Br$\gamma$, and \ion{He}{I} at 1.70\,$\mu$m but a weak Brackett series. This is characteristic 
of late O- to early B-dwarfs. Comparing with spectra from the Hanson catalogues, we 
find that the spectral type of this object corresponds to B2\,V. 

 In the case of stars \#6 ($H=13.75$ and $K_S=12.66$) and \#7 ($H=14.03$ and $K_S=13.24$), 
it was impossible to derive a precise spectral type owing to the presence of some emission
 lines, the lack of other strong spectral features, and the low $SNR$ of our data for these dim objects.

 In the centre of the cluster, nebular emission is an important contributor to the spectra 
and, in some objects, appears to be inhomogeneous. Star \#8, practically in the middle 
of the field, has been resolved thanks to our LIRIS image resolution. The nebular emission 
is overcorrected because of the inhomogeneities in the sky emission around the object. 
The hydrogen lines exhibit a wide emission component mixed with shallower
absorption. This might be due to the overlapping of the nebular (wide) contribution 
with stellar (narrow) lines. We also detect \ion{He}{I} at 1.7 and {2.06\,$\mu$m for this object. 
The position of the star \#8 in the colour-colour diagram, below the main-sequence line, 
indicates that it is a YSO-like star \#2. Because it is less reddened, it might have cleared 
its surroundings.

 A similar effect could occur for star \#9; its spectra looks featureless in absorption,
probably because its lines are filled by the nebular emission lines. The classification 
of the spectrum, dominated by Brackett and \ion{He}{I} emission lines, was impossible.

 Star \#10 has a well-defined Brackett series, with no indication of either helium or carbon
lines. A comparison with star HR5685 from Ranade's library leads us to classify this object
as B5\,V.

 Stars \#11 and \#12 are late-type stars, even if the CO bands do not clearly appear in 
these spectra. A close similarity can be found between star \#11 and Ranade's HR5019 
(spectral type G6.5\,V) and between star \#12 and Meyer's HR4496 (spectral type G8 V). 
Both appear to be stars in the Galactic disc.

  Finally, star \#13 has a remarkable position in the colour--magnitude diagram, being 
the brightest star in our selection. However the weak Brackett and
the strongest CO heads in our candidates indicate a late-type for this star.
 The spectra looks similar to those of HR5888 \citep{ivanov04} and HR7328 
 \citep{meyer98}. We classify star \#13 as G8-G9\,III.

\subsection{Spectral classification (optical)}
 Our final optical spectrum for the central star of Sh2-152, is shown in Fig. \ref{espectroptico}. 
This star was previously classified as O9\,V by \citet{crampton78} and O8.5\,V by \citet{russeil07}. 
 
 Our spectral classification is based on the study of \citet{walborn90}. On first inspection, 
the spectral type for the central star of Sh2-152 appears to be O9-B0\,V star, because of
its ratio of \ion{He}{I} 4471 \AA~to \ion{He}{II} 4542 \AA. However, there are a few spectral
features in the optical spectra that deserve some review. 

 First of all, the \ion{He}{II} 4686 \AA~line is similar in depth to the \ion{He}{I} 4471 \AA~line 
and deeper than the \ion{He}{I} 4388 \AA~one. The ratio \ion{He}{II} 4686 \AA$-$\ion{He}{I} 
4471 \AA~looks slightly larger than for stars between O9\,V and B0\,V. An \ion{He}{II} 4686 
\AA~spectral line stronger than that centred on \ion{He}{I} 4471 \AA~would indicate a Vz 
type \citep{walborn09} and might indicate that there is ongoing massive-star formation in 
the cluster. 
 
 Secondly, the CNO complex at 4630-4650 \AA~looks slightly fainter than expected for those
spectral types. Although this might be caused by the continuum rectification, this would not 
seem to be sufficient to justify the difference.

 The third interesting feature is the absence of a \ion{Si}{III} (4553--4574 \AA) detection. 
These lines appear for spectral types later than O9.5\,V, hence this detection indicates a
spectral type of O9\,V or earlier. However, for types earlier than the aforementioned limit, 
the depths of the \ion{He}{II} 4542 \AA~and \ion{He}{II} 4686 \AA~lines do not fit.
 
  Therefore, the star was assigned a spectral type of O9$-$B0\,V, with spectral features 
that are not fully consistent with any of the subtypes. While the $SNR$ of the spectrum 
is relatively modest (the $SNR$ per pixel is 20, but at $R=46000$, which allows us to degrade
our spectrum without loss of information), we cannot exclude the presence of a binary system,
a composite spectrum, or a slightly peculiar spectrum. The possibility of a Vz type would 
also enhance the peculiarity of this star, but data of higher $SNR$ is mandatory to confirm this point.

\begin{figure} 
\centering
\includegraphics[width=2.85in,angle=270]{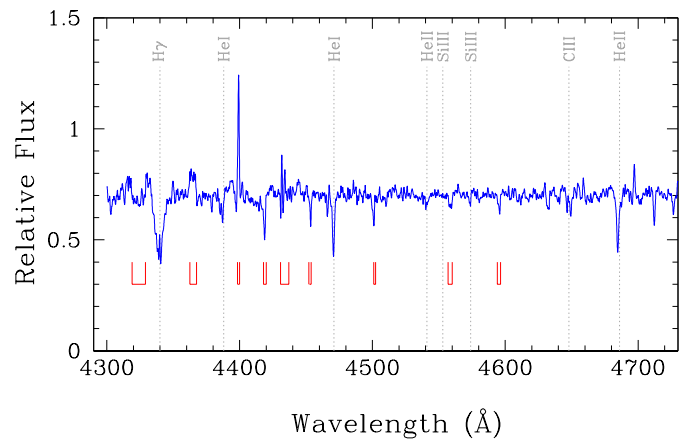}
            \caption{Optical spectrum for the central star of Sh2-152 (Russeil's star number 4). Vertical segmented 
            lines (grey) mark the spectral lines used in the spectral classification. Overlapping zones of the spectral 
            orders are marked with red vertical lines; the low $SNR$ causes a poor combination of the spectral 
            orders, which could mimic a spectral line. Spectra resolution was decreased to $R=8000$, to help us
            to discern the lines.}
       \label{espectroptico}
    \end{figure}


 \section{Discussion}
  \subsection{Distance and size determinations}

 Using the estimated spectral types, we derived the individual distances to cluster
stars. Assuming the absolute visual magnitudes of \citet{cox00}, the intrinsic infrared 
colours of \citet{tokunaga00}, and the \citet{cardelli89} extinction law with $R=3.10$,
 the extinction for the $K_S$ band can be expressed as

\begin{equation}
A_{K_S} = \frac{E_{J-K_S}}{1.474} = \frac{E_{H-K_S}}{0.667}
\end{equation}
Across the field of view, we can find several situations of extinction ranging from 
0.62 to 2.70 mag for $A_{K_S}$, or $A_V$ between 5.44 and 23.68 mag. In Table \ref{distances}, 
we present the values of $A_{K_S}$ and the individual distances of the early-type stars 
used in the distance estimation.
  
   We excluded Russeil's star number 4 (central star of Sh2-152) because it appears saturated
in our images, and 2MASS images of this object are unable to resolve the central zone of the
cluster. The 2MASS magnitudes for the central star would be contaminated and we prefer to
exclude this object from the distance estimation.

\begin{table}
\caption{Reddening and distances of early-type stars.}
\begin{center}
\begin{tabular}{cccccc}
\toprule
 Spectral type & ID & $A_{K_S}$ & Distance \\
 &  & [mag] & [kpc] & &\\
 \midrule
B1 V	         & 1   &  2.70 & 3.17 \\
B2 V          & 5   &  0.70 & 3.76 \\
B2--3 V	& 3   &  0.76 & 3.61 \\
B2--3 V	& 4   &  0.62 & 2.70 \\
B5 V          & 10 &  0.95 & 2.83 \\
\bottomrule
\end{tabular}
\end{center}
\label{distances}
\end{table}

 With the five early-type stars, we obtained an average distance to the cluster of
$3.21\pm0.21$ kpc from the individual calculations. This value is larger than that
 given by \citet{russeil07} - $D=2.39$ kpc- and slightly smaller than the distance of 3.5 kpc 
estimated by \citet{crampton78}. Both estimates were obtained from spectral observations 
of the central ionizing star of the cluster (our O9$-$B0\,V star). In our case, the estimate
 was derived from individual distance estimates that are self-consistent and cover a
 wider area of the cluster.

  From the Spitzer image of the field and our observation of a nearby control field
 using LIRIS, we learned that the angular size of the cluster is 
comparable to the LIRIS field of view ($4.2\arcmin \times 4.2\arcmin$). The cluster was previously
 considered as just the surroundings of Sh2-152, underestimating the size of the object, 
but from our CMD, we realize that stars near the edge of our image are part of the cluster. 
Hence, considering at least a cluster size of $4.2'$ (the LIRIS field of view) and the average 
distance of 3.21 kpc, the diameter of the cluster would be $\gtrsim4.0$ pc.

 A test that helps us to clarify the extension of the cluster is to count the number of stars 
for different concentric rings covering the whole field of view. Taking as centre 
the coordinates of IRAS 22566+5830, we counted the sources in concentric rings of 25 
arcsec in thickness. In Fig. \ref{radialcounts}, we can see that there is an evident change in 
the slope of the counts in the central section of the image (around 250 pixels, i.e.\ 60
 arcsec), but even as the number of stars detected continues to decrease it remains larger 
than the number of stars detected for the control field. In addition, the existence of stars with 
the same spectrophotometric distances as objects within the central area of the cluster
 allows us to confirm that the size of the cluster is at least comparable to our field of
 view, and that the subtraction of stellar counts using the same field is unreliable.

\begin{figure} 
\centering
\includegraphics[width=2.5in,angle=270]{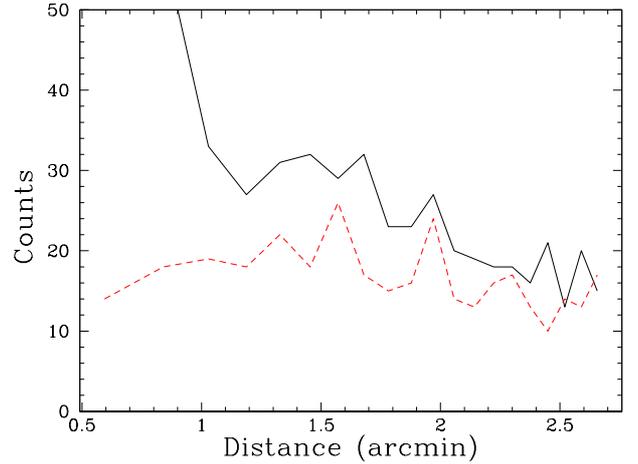}
            \caption{Radial distribution of stars for Sh2-152 (solid black curve) and 
the control field (red dashed curve).}
       \label{radialcounts}
    \end{figure}

\subsection{Cluster mass and age estimations}
\label{sec:clustermass}

 To derive the cluster mass, we integrated the stellar mass function of Sh2-152, 
previously corrected for the field stellar contribution, after accounting for the effect of 
completeness on the star counts.
 
 The first step in obtaining the mass function is to project every star following the reddening
vector, to the dwarf star sequence defined by the magnitudes and colours of \citet{cox00} 
and located at our estimated distance. This sequence can be expressed analytically by a
second degree function or a set of first degree functions. We found that fitting the dwarf 
sequence using two lines, one from O9\,V to A0\,V and the second from A0\,V to F0\,V 
(blue solid lines in Fig. \ref{id436_CMDs}) was more effective than fitting a second-degree
function. The cut in F0\,V arises from our detection limit; at later spectral types (fainter 
magnitudes), the histograms start to decline towards lower masses.
 
 Once the stars in the Sh2-152 and the control field CMDs were projected following the 
reddening vector to the line-fitted main sequence, we produced the histograms including
the specific weight for each star, according to its completeness factor determined by its 
position in the CMD ($J-K_S$, $K_S$). The correction for the incompleteness consists 
simply of multiplying the number of stars by the inverse of the completeness factor.

 The histograms bins were 0.8 in $K_S$, equivalent to twice the size of the completeness
bins. To convert a $K_S$ magnitude to a stellar mass, we used the values given by 
\citet{cox00}. For in-between magnitudes catalogued values, we interpolated between
the two closest values.

 After subtracting both histograms (i.e. subtracting the control field histograms from the
Sh2-152 field), we derived the cluster mass function (Fig. \ref{histograms}). Deriving the
best-fit linear relation ($\Gamma = -1.62$) in the log(N)--log(M) plane and then integrating
from 0.20 to 20 ${M}_{\odot}$, we obtained a total mass for Sh2-152  of $(2.45\pm0.79)\cdot10^3 {M}_{\odot}$. 

\begin{figure}
\centering
\includegraphics[width=3.5in]{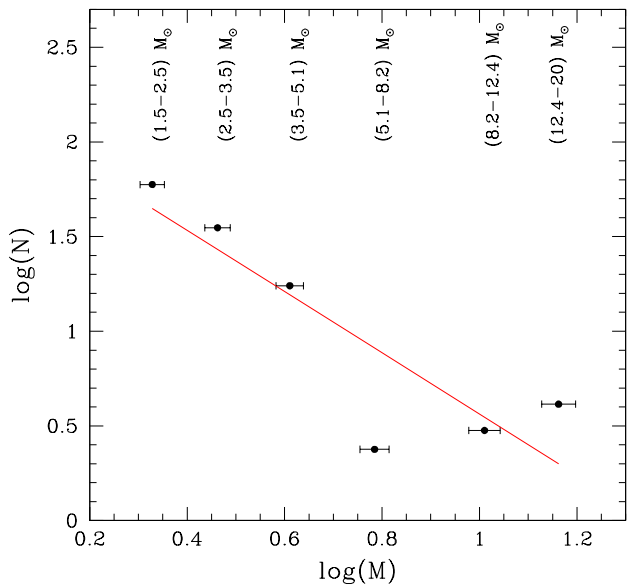}
            \caption{Mass function for Sh2-152. The points indicate the central position of the
            mass range indicated above, equivalent to $K_S$ magnitude bins of 0.8. The 
            continuous red line shows the best linear fit to the data.}
       \label{histograms}
    \end{figure}

 To compare this result with the literature, we had to carefully consider the nature of both
quantities. Our value comes from evaluating the stellar content of Sh2-152, in sharp contrast
to studies such as \citet{ao04}, \citet{guan08}, and \citet{curran04}, who derived the virial mass, 
i.e.\ the stellar mass necessary to ensure that the molecular core is gravitationally bound, from
molecular radial velocities.
 
 \citet{guan08} calculated} the mass (dust) for the associated CO core. Their value 
for the total mass (dust) is $3.81\cdot10^4 {M}_{\odot}$, an order of magnitude 
higher than our result. If we use our distance estimate instead of the value used by
 \citet{guan08} (5.3 kpc), the total mass (dust) for the CO core drops to
 $1.22\cdot10^4 {M}_{\odot}$ and its virial mass is also reduced by a factor of 
0.32, down to  $3.04\cdot10^3 {M}_{\odot}$.

 The study of \citet{ao04} assumes a distance that is more similar to our estimate. 
With $d=3.5$ kpc, they obtained four total mass (dust) estimates: $1.78\cdot10^3 {M}_{\odot}$ 
for the CS core, $9.20\cdot10^3 {M}_{\odot}$ for \element[][13]{CO} core, and 
$(0.29-1.20)\cdot10^4 {M}_{\odot}$ for the two C\element[][18]{O} cores. With 
the exception of the virial mass derived from the \element[][13]{CO} estimate 
$(4.90\cdot10^3 {M}_{\odot}$), their associated virial masses are lower 
than those obtained by \citeauthor{guan08}: $1.78\cdot10^3 {M}_{\odot}$ 
(for CS core) and $(0.67-1.40)\cdot10^3 {M}_{\odot}$ (for the C\element[][18]{O} cores).
  
 With the stellar mass and dust mass estimates, we can evaluate the cluster's star
formation efficiency (SFE). The star formation efficiency is the ratio of the cluster's stellar
mass to the gas-stellar mass. To derive the total gas mass associated with the dust total
mass, we used the gas-to-dust ratio. According to \citet{curran04}, this ratio varies from 
45:1 to 100:1, so we prefer to use both values in the estimate.
  
  Using  the dust mass measured by \citet{curran04}, projected to our estimated distance
instead of the 5.0 kpc used by them, we obtained the associated gas mass, and estimated
the SFE to be between 1.7\% and 3.7\%. This value is similar to that reported by \citet{hunter90} (i.e. 6\%), 
 assuming a distance of 4.0 kpc and estimating the stellar total mass with a normalization
 of the IMF to the upper stellar range, which in their case is $18-25 {M}_{\odot}$.
  
 We can neither precisely determine the age of the cluster nor confirm the idea of sequential
stellar evolution \citep{chen09}, but we can constrain the cluster's age based on the 
physical characteristics of its stellar population. First of all, our earliest main-sequence star
is of type O9\,V. Since it has not evolved away from the main sequence, the cluster lifetime
is shorter than the time spent by this star on the main sequence. Assuming that $18\,{M}_{\odot}$
is the mass of an O9\,V star \citep{martins05} and that the cluster population was formed at
the same time, we can only limit the cluster's age to be younger than 9.4 Myr, using the stellar
models for a $20\,{M}_{\odot}$ solar metallicity star of \citet{schaller92}. This limit is obviously
conservative and cannot be considered an age estimate.

\subsection{Ionizing sources}
 
 We used our spectral type classification together with data collected from the literature to 
estimate whether the total ionizing photons emitted by the stars can explain the radio observations 
of Sh2-152. To determine the number of ionizing photons (log\,$Q_0$) for O-type stars, we used
\citet{martins05}. For B0\,V stars, we used the values given by \citet{simondiaz08} and for the 
B0.5$-$B1\,V spectral types we took the radii from the binary CW Cep \citep{bozkurt07}, the 
spectral type$-T_\mathrm{eff}$ relation from \citet{lefever10}, and the $T_\mathrm{eff}-\mathrm{log}\,q_0$
from the B-star grid from \citet{lanz07}.
 
 To derive the number of Lyman continuum photons from the radio density flux, $N_{\mathrm LyC}$, 
we used Eq. A.5 of \citet{hunt04}, assuming an electron temperature of $T=10^4 \mathrm{K}$ and
$n\mathrm{(He^+)} \ll n\mathrm{(H^+)}$. Since this formula is for radiation-bounded ionized \ion{H}{II} 
 regions, it is not guaranteed that it is applicable to IRAS 22566+5828, although an estimation may be illustrative.
 
 We divide the ionizing stars into two groups: the first contains stars in the central region ($r\sim30\arcsec$), 
including IRAS 22566+5830, star \#10 and Russeil's star number 4. The earliest star in this region is Russeil's 
number 4 (O9$-$B0\,V, this work), with a total number of ionizing photons between log\,$Q_0 = 48.06$ and 
47.29. The number of Lyman continuum photons derived from the radio observations for IRAS 22566+5830 is 
log\,$N_{\mathrm LyC}= 48.02$ \citep{condon98}, which is most consistent with an O9\,V spectral classification
for the central star of Sh2-152.
 
  The second region, formed by stars \#1, \#3, \#4, and \#5, is located close to IRAS 22566+5828 source
and its earliest star is star \#1. Within our sample, star \#1 would therefore be the main ionizing source
associated with IRAS 22566+5828. However, the value of log\,$Q_0 = 45.66$ for this star is significantly
lower than for the Lyman continuum photons, log\,$N_{\mathrm LyC}= 48.35$, derived from the radio 
observations \citep{harju98} indicating that the star \#1 does not produce enough photons to explain the radio 
observations. The $K_S$ image shows a deeply embedded star in the centre of the red nebulosity close to
of IRAS 22566+5828, which appears to be the O7.5$-$8\,V star required as the single ionizing source of 
this region. Future $K$ spectroscopy for this star would help us to confirm this result.


\section{Conclusions}

 In the framework of our MASGOMAS project, we have completed a spectrophotometric study of 
the stellar content of the compact \ion{H}{II} region Sh2-152 using LIRIS at the WHT and the
 FIES cross-disperser echelle spectrograph at the NOT.

 Our new near-infrared data have allowed us to resolve the central part of the cluster, to select 
OB candidates for $H$ and $K$ spectroscopy, and estimate their individual distances. The mean
 of the individual distances to five B-type stars, $3.21\pm0.21$ kpc, places the cluster slightly closer 
than the estimates obtained by \citet{crampton78, wouterloot86, wouterloot88}, and \citet{harju93}
but definitely further away than estimated by \citet{russeil07}. In any case, our estimated distance 
is in closer agreement with the distance estimates based on the central ionizing star (star number 
4 in \citealt{russeil07}) than with those derived from \element[][12]{CO} radial velocities.

  We also obtained the cluster mass function, corrected for the Galactic disc stellar 
contribution (using the observed control field) and for completeness effects (through 
artificial star injection and photometry). Integrating this function, we calculated
 a lower limit to the cluster total mass. The value of $(2.45\pm0.79)\cdot10^3 
{M}_{\odot}$ agrees with the mass values from the literature, all of them being 
virial masses necessary to reproduce the observed molecular velocities (for both 
CO and CS cores).

The central O9--B0\,V star remaining on the main sequence allows us to limit the 
cluster's age to be younger than 9.4 Myr. The optical spectra for this central star displays
several interesting features. First, the strenght of its helium lines (\ion{He}{II} 4686 \AA~similar to
that of \ion{He}{I} 4471 \AA~and stronger than \ion{He}{I} 4388 \AA) might be indicative 
of a V$z$ spectral type, although the relation between the \ion{He}{II} 4686 \AA~line and
\ion{He}{II} 4542 \AA~line appears to exclude this. Secondly, the spectra detect a fainter CNO
complex at 4630--4650 \AA~for the derived spectral type. And third, we have found the absence of 
\ion{Si}{III} (4553 \AA~and 4574 \AA). Either a binary system, a composite spectrum, or a 
slightly peculiar spectrum are possible explanations of the last two features. Future 
observations of this object with higher $SNR$ will help us to shed light on the real 
nature of the central, most massive star of Sh2-152.
 
 For two regions of Sh2-152, we have compared the number of ionizing photons from the classified 
OB-stars with the number of Lyman continuum photons derived from radio observations. For the
central region containing both IRAS 22566+5830 and the O9$-$B0\,V central star (Russeil's star 
number 4), both of these numbers are consistent with there being a single ionizing source.

 For the second region containing IRAS 22566+5828, the number of Lyman continuum
photons cannot be explained by the earliest star in our sample, and the derived value of 
log\,$N_{\mathrm LyC}$ would imply the existence of an O7.5$-$8\,V star. This source could
be a highly embedded star, without spectral classification, at the centre of the bright red arc 
close to star \#1. We expect future spectral observations of this star to help us verify this finding.

 
\begin{acknowledgements}

S.R.A. was supported by the MAEC-AECID scholarship. Part of this work was 
supported by the Science and Technology Ministry
of the Kingdom of Spain (grants AYA2008-06166-C03-01), and the Fundaci\'on 
Agencia Aragonesa para la Investigaci\'on y Desarrollo (ARAID). S.R.A., A.H., A.M-F., 
E.P., F.N.,  J.A.A.P., and S.S.-D. are members of the Consolider-Ingenio 2010 Program (CSD2006-00070).

The William Herschel Telescope is operated on the island of La Palma by the 
Isaac Newton Group in the Spanish Observatorio del
Roque de los Muchachos of the Instituto de Astrof\'isica de Canarias. This
 publication makes use of data products from the Two Micron All Sky Survey, 
which is a joint project of the University of Massachusetts and the Infrared 
Processing and Analysis Center/California Institute of Technology, funded by 
the National Aeronautics and Space Administration and the National Science Foundation.

\end{acknowledgements}

\bibpunct{(}{)}{;}{a}{}{,} 
\bibliographystyle{aa} 
\bibliography{biblio}

\end{document}